\documentclass[12pt,preprint]{aastex}

\slugcomment{}

\shorttitle{Two New Lensed Quasars from the SDSS}
\shortauthors{Andi et ala.}

\begin{document}

\title{SDSS~J0806+2006 and SDSS~J1353+1138: Two New Gravitationally 
Lensed Quasars from the Sloan Digital Sky Survey}

\author{
Naohisa Inada,\altaffilmark{1}
Masamune Oguri,\altaffilmark{2,3}
Robert H. Becker,\altaffilmark{4,5}
Richard L. White,\altaffilmark{6}
Michael D. Gregg,\altaffilmark{4,5}
Paul L. Schechter,\altaffilmark{7}
Yozo Kawano,\altaffilmark{8} 
Christopher S. Kochanek,\altaffilmark{9} 
Gordon T. Richards,\altaffilmark{10}
Donald P. Schneider,\altaffilmark{11}
J. C. Barentine,\altaffilmark{12}
Howard J. Brewington,\altaffilmark{12}
J. Brinkmann,\altaffilmark{12}
Michael Harvanek,\altaffilmark{12}
S. J. Kleinman,\altaffilmark{12}
Jurek Krzesinski,\altaffilmark{12,13}
Dan Long,\altaffilmark{12}
Eric H. Neilsen, Jr.,\altaffilmark{14}
Atsuko Nitta,\altaffilmark{12}
Stephanie A. Snedden,\altaffilmark{12}
and Donald G. York\altaffilmark{15,16} 
}

\altaffiltext{1}{Institute of Astronomy, Faculty of Science, 
University of Tokyo, 2-21-1 Osawa, Mitaka, Tokyo 181-0015, Japan.}
\altaffiltext{2}{Princeton University Observatory, Peyton Hall,
  Princeton, NJ 08544.} 
\altaffiltext{3}{Department of Physics, University of Tokyo, Hongo
  7-3-1, Bunkyo-ku, Tokyo 113-0033, Japan.} 
\altaffiltext{4}{IGPP-LLNL, L-413, 7000 East Avenue, Livermore, CA 94550.}
\altaffiltext{5}{Physics Department, University of California, Davis,
  CA 95616.} 
\altaffiltext{6}{Space Telescope Science Institute, 3700 San Martin
  Drive, Baltimore, MD 21218.} 
\altaffiltext{7}{MIT Kavli Institute for Astrophysics and Space Research, 
  77 Massachusetts Avenue, Cambridge MA 02139.} 
\altaffiltext{8}{Department of Physics and Astrophysics, Nagoya
  University, Chikusa-ku, Nagoya 464-8062, Japan.} 
\altaffiltext{9}{Department of Astronomy, The Ohio State University, 
  Columbus, OH 43210.}
\altaffiltext{10}{Department of Physics and Astronomy, The Johns Hopkins
University, 3400 North Charles Street, Baltimore, MD 21218-2686.}  
\altaffiltext{11}{Department of Astronomy and Astrophysics, The
  Pennsylvania State University, 525 Davey Laboratory, University
  Park, PA 16802.}  
\altaffiltext{12}{Apache Point Observatory, P.O. Box 59, Sunspot, NM 88349.}
\altaffiltext{13}{Mt. Suhora Observatory, Cracow Pedagogical University, ul. Podchorazych 2,
  30-084 Cracow, Poland.}
\altaffiltext{14}{Fermi National Accelerator Laboratory, P.O. Box 500, Batavia, IL 60510.}
\altaffiltext{15}{Department of Astronomy and Astrophysics, The University of Chicago, 
  5640 South Ellis Avenue, Chicago, IL 60637.}
\altaffiltext{16}{Enrico Fermi Institute, The University of Chicago, 5640 South 
  Ellis Avenue, Chicago, IL 60637.}
  
\begin{abstract}
We report the discoveries of two, two-image gravitationally lensed quasars 
selected from the Sloan Digital Sky Survey: SDSS~J0806+2006 at $z_s=1.540$
and SDSS~J1353+1138 at $z_s=1.629$ with image separations of 
$\Delta{\theta}=1\farcs40$ and $\Delta{\theta}=1\farcs41$ 
respectively.  Spectroscopic and optical/near-infrared imaging follow-up
observations show that the quasar images have identical redshifts and 
possess extended objects between the images that are likely to be 
lens galaxies at $z_l \simeq 0.6$ in  SDSS~J0806+2006 
and $z_l \simeq 0.3$ in SDSS~J1353+1138.  The field of SDSS~J0806+2006 
contains several nearby galaxies that may significantly perturb the system,
and  SDSS~J1353+1138 has an extra component near its Einstein ring
that is probably a foreground star. Simple mass models with reasonable 
parameters reproduce the quasar positions and fluxes of both systems.
\end{abstract}


%%%%%%%%%%%%%%%%%%%%%%%%%%%%%%%%%%%%%%%%%%%%%%%%
%%%%%%%%%%%%%%%%%%%%%%%%%%%%%%%%%%%%%%%%%%%%%%%%
%%%%%%%%%%%%%%%%%%%%%%%%%%%%%%%%%%%%%%%%%%%%%%%%
\section{Introduction}\label{sec:intro}
%%%%%%%%%%%%%%%%%%%%%%%%%%%%%%%%%%%%%%%%%%%%%%%%
%%%%%%%%%%%%%%%%%%%%%%%%%%%%%%%%%%%%%%%%%%%%%%%%
%%%%%%%%%%%%%%%%%%%%%%%%%%%%%%%%%%%%%%%%%%%%%%%%

Since the discovery of Q0957+561 \citep*{walsh79}, about 80
gravitationally lensed quasars have been discovered
\citep{kochanek04}. Lensed quasars are not only intriguing phenomena but
also have become indispensable astronomical tools, including probes of
the cosmological parameters and the structure of galaxies
\citep[e.g.,][]{refsdal64,kochanek91}. In particular, the
abundance of gravitational lenses in a well-defined source sample
can be used to constrain dark energy \citep[e.g.,][]{turner90,fukugita90,chae02}. 
Unfortunately, the largest existing survey, the Cosmic Lens
All-Sky Survey \citep[CLASS;][]{myers03,browne03}, contains only 22
lensed radio sources (with a well-defined statistical sample of 13
lenses) discovered from  ${\sim}10,000$ radio sources, 
which is still insufficient to place tight constraints on dark energy
models. The Sloan Digital
Sky Survey \citep[SDSS;][]{york00} should lead to a significantly
larger lens sample for attacking the dark energy problem.  SDSS
is expected to identify $10^5$ quasars spectroscopically \citep[e.g.,][]{schneider05} 
and $\approx10^6$ quasars photometrically \citep[e.g.,][]{richards04}, which
should lead to a sample of over $10^2$ lensed quasars given a standard 
lensing probability of $10^{-3}$ \citep{turner84}.  Indeed, 12 new lensed 
quasars have been discovered from the SDSS quasars so far 
\citep*{inada03a,inada03b,inada03c,inada05,johnston03,morgan03,pindor04,
pindor05,oguri04,oguri05,burles05}, in addition to recovering 5
previously known lensed quasars \citep{walsh79,weymann80,surdej87,bade97,oscoz97}. 
We can presently construct a well-defined statistical sample of 16 lensed SDSS 
quasars, but there remain many promising SDSS lensed quasar candidates for 
which the necessary follow-up observations are incomplete.

In this paper, we report on the discovery of two more
gravitationally lensed quasars, SDSS~J080623.70+200631.9
(SDSS~J0806+2006) and SDSS~J135306.35+113804.7 (SDSS~J1353+1138). We
present imaging and spectroscopic follow-up observations with the
University of Hawaii 2.2-meter (UH88) telescope, the W. M. Keck
Observatory's Keck I and II telescopes, and the Magellan Consortium's
Landon Clay 6.5-m (LC6.5m) telescope \footnote{The second telescope of
the Magellan Project; a collaboration between the Observatories of
the Carnegie Institution of Washington (OCIW), University of
Arizona, Harvard University, University of Michigan, and
Massachusetts Institute of Technology (MIT) to construct two 6.5 Meter
optical telescopes in the southern hemisphere.}. We model the systems 
to check that their geometries are consistent with the lensing 
hypothesis.

The structure of this paper is as follows. We describe our lens
candidate selections from the SDSS data in \S \ref{sec:sdss}.
The follow-up observations and mass modeling of SDSS~J0806+2006 
and  SDSS~J1353+1138 are presented in \S \ref{sec:0806} and 
\S \ref{sec:1353}, respectively. We finally present a summary and give
a conclusion in \S \ref{sec:conc}.  Throughout the paper 
we assume a cosmological model with the matter
density $\Omega_M=0.27$, cosmological constant $\Omega_\Lambda=0.73$,
and Hubble constant $h=H_0/100{\rm km\,sec^{-1}Mpc^{-1}}=0.7$
\citep{spergel03}.

%%%%%%%%%%%%%%%%%%%%%%%%%%%%%%%%%%%%%%%%%%%%%%%%
%%%%%%%%%%%%%%%%%%%%%%%%%%%%%%%%%%%%%%%%%%%%%%%%
%%%%%%%%%%%%%%%%%%%%%%%%%%%%%%%%%%%%%%%%%%%%%%%%
\section{Selecting Lensed Quasar Candidates from the SDSS}\label{sec:sdss}
%%%%%%%%%%%%%%%%%%%%%%%%%%%%%%%%%%%%%%%%%%%%%%%%
%%%%%%%%%%%%%%%%%%%%%%%%%%%%%%%%%%%%%%%%%%%%%%%%
%%%%%%%%%%%%%%%%%%%%%%%%%%%%%%%%%%%%%%%%%%%%%%%%

The SDSS is conducting a photometric and spectroscopic survey of 10,000
square degrees of the sky approximately centered  on the North Galactic
Cap using the dedicated wide-field ($3^{\circ}$  field of view) 2.5-m
telescope \citep{gunn05} at the Apache Point Observatory in New Mexico,
USA. Photometric observations \citep*{gunn98,lupton99,tucker05} are made 
in five optical filters \citep{fukugita96}. After automated data processing 
by the photometric pipeline \citep{lupton01,lupton05}, quasar and galaxy 
candidates are selected by the spectroscopic target selection algorithms 
\citep{eisenstein01,richards02,strauss02}.
Spectra of these candidates are obtained according to the tiling algorithm 
of \citep{blanton03} using a multi-fiber spectrograph covering
3800{\,\AA} to 9200{\,\AA} at a resolution of R$\sim1800$. 
The data are very homogeneous, with an astrometric accuracy better than
about $0\farcs1$ rms per coordinate \citep{pier03} and photometric zeropoint
errors less than about 0.03 magnitude over the entire 
survey area \citep{hogg01,smith02,ivezic04}. The data have been released
continuously to the public
\citep{stoughton02,abazajian03,abazajian04,abazajian05,adelman05}. 

We selected the two objects, SDSS~J0806+2006 and SDSS~J1353+1138, as 
lensed quasar candidates from the SDSS spectroscopic quasar sample 
with the same algorithm used for the discovery of most SDSS lensed
quasars \citep[e.g.,][]{inada03a}.  Specifically, the algorithm uses the
SDSS image parameters {\tt dev\_L}, {\tt exp\_L} and {\tt star\_L} for
the likelihood that a source can be modeled as a de Vaucouleurs 
profile, an exponential disk or a point source to quantify the
structure of each quasar.  Lensed quasars should be modeled poorly 
by all three profiles, so quasars with small values for all three
likelihoods are good lens candidates \citep*[see][for more details]{inada03a}. 

\begin{figure}
\epsscale{.46}
\plotone{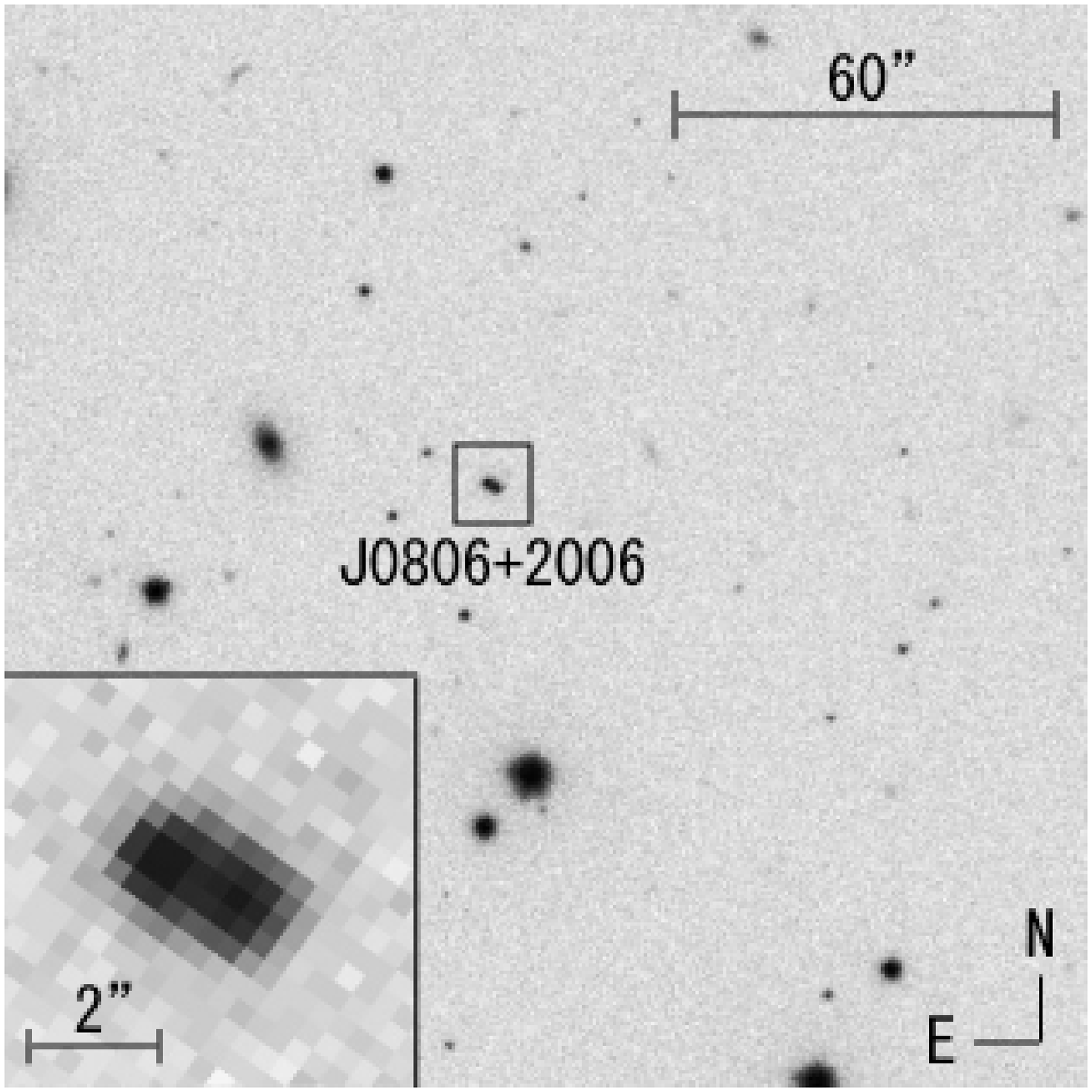}
\plotone{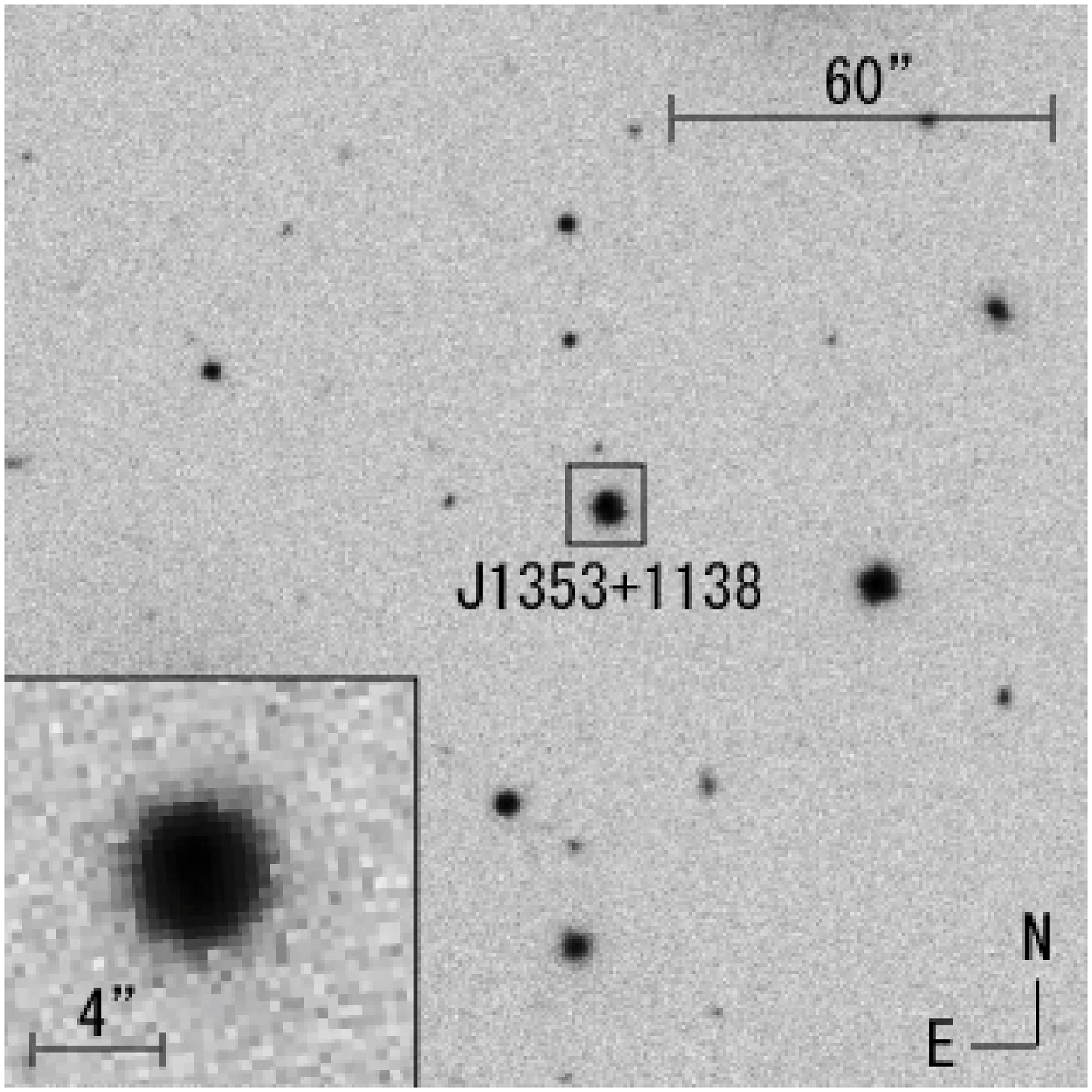}
\caption{{\it Left:} The SDSS $i$-band image of the field centered on 
 SDSS~J0806+2006. The image scale is 0\farcs396 ${\rm pixel^{-1}}$. 
 The enlarged SDSS~J0806+2006 image is shown in the inset. In
 both the field image and the inset, North is up and East is left.
 {\it Right:} Same as the left panel, but for SDSS~J1353+1138. 
\label{fig:field08061353}}
\end{figure}

The SDSS $i$-band images of SDSS~J0806+2006 and SDSS~J1353+1138
are shown in Figure~\ref{fig:field08061353}. Both systems are clearly
resolved, making them excellent lens candidates since their spectra
are clearly those of $z>1$ quasars.
For SDSS~J0806+2006, the total magnitudes
(within ${\sim}2\farcs3$ aperture radius) of the system are
$19.18\pm0.05$, $18.76\pm0.01$, $18.44\pm0.02$, $18.06\pm0.02$, and
$17.92\pm0.05$, in $u$, $g$, $r$, $i$, and $z$, respectively. 
For SDSS~J1353+1138, the total magnitudes ($ugriz$, within 
${\sim}2\farcs1$ aperture radius) are $16.87\pm0.01$, $16.60\pm0.01$, 
$16.48\pm0.01$, $16.26\pm0.01$, and $16.20\pm0.02$, respectively. 
The redshifts of SDSS~J0806+2006 and SDSS~J1353+1138 measured from their 
SDSS spectra are $z_s=1.537\pm0.002$ and $z_s=1.623\pm0.003$, respectively. 
From the SDSS spectra, with a fiber aperture of $3\farcs0$, we know that
the components of the candidates cannot have greatly dissimilar
spectra. However, we cannot resolve the spectra of the individual
components. 

Thus, while the SDSS data are sufficient to identify these objects as lens
candidates, additional observations are needed to confirm them as lensed quasars.
Deeper and higher resolution images are needed to confirm the existence 
of multiple quasar images and to search for the lens galaxies, 
and spatially resolved spectra are needed to confirm that the quasar images 
have identical redshifts. We present this evidence for SDSS~J0806+2006 in
\S3 and for SDSS~J1353+1138 in \S4.

%%%%%%%%%%%%%%%%%%%%%%%%%%%%%%%%%%%%%%%%%%%%%%%%
%%%%%%%%%%%%%%%%%%%%%%%%%%%%%%%%%%%%%%%%%%%%%%%%
%%%%%%%%%%%%%%%%%%%%%%%%%%%%%%%%%%%%%%%%%%%%%%%%
\section{SDSS~J0806+2006}\label{sec:0806}
%%%%%%%%%%%%%%%%%%%%%%%%%%%%%%%%%%%%%%%%%%%%%%%%
%%%%%%%%%%%%%%%%%%%%%%%%%%%%%%%%%%%%%%%%%%%%%%%%
%%%%%%%%%%%%%%%%%%%%%%%%%%%%%%%%%%%%%%%%%%%%%%%%

%%%%%%%%%%%%%%%%%%%%%%%%%%%%%%%%%%%%%%%%%%%%%%%%
\subsection{Imaging Observations}\label{sec:0806img}
%%%%%%%%%%%%%%%%%%%%%%%%%%%%%%%%%%%%%%%%%%%%%%%%

\begin{figure}
\epsscale{.95}
\plotone{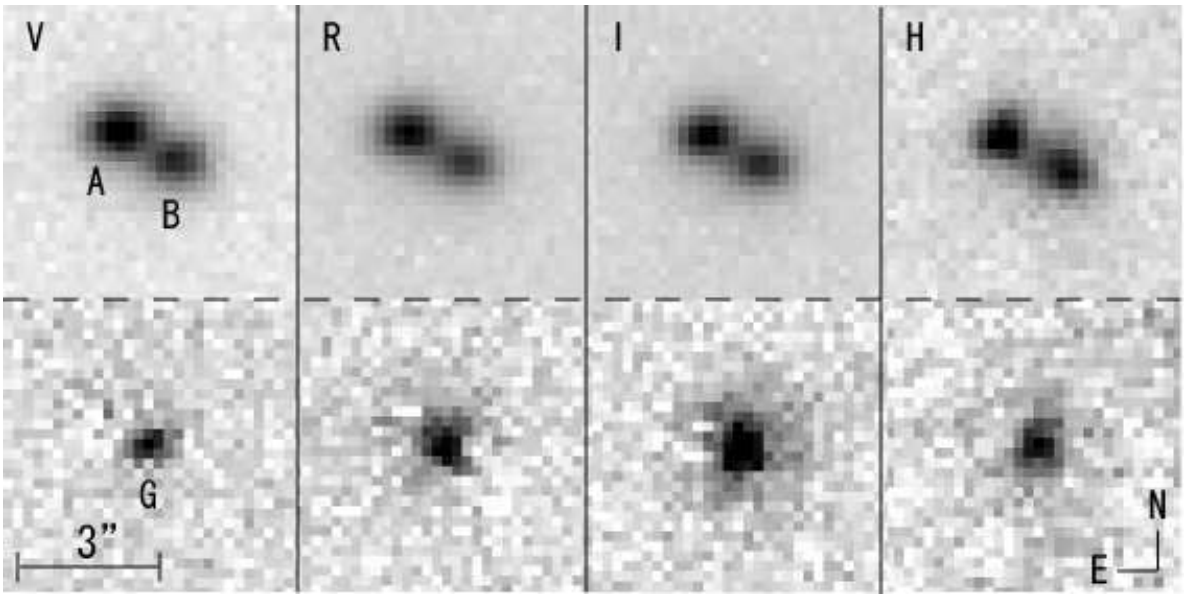}
\caption{The $VRIH$-band images of SDSS~J0806+2006 taken with the 8k
 mosaic CCD camera and QUIRC at the UH88. The image scales are 
 0\farcs235 ${\rm pixel^{-1}}$ for the 8k mosaic CCD camera and 0\farcs189 ${\rm
 pixel^{-1}}$ for QUIRC.  The lower panels show the images after
 subtracting components A and B: The images clearly show the lens
 galaxy (component G). 
\label{fig:uh88_0806}}
\end{figure}

\begin{figure}
\epsscale{.7}
\plotone{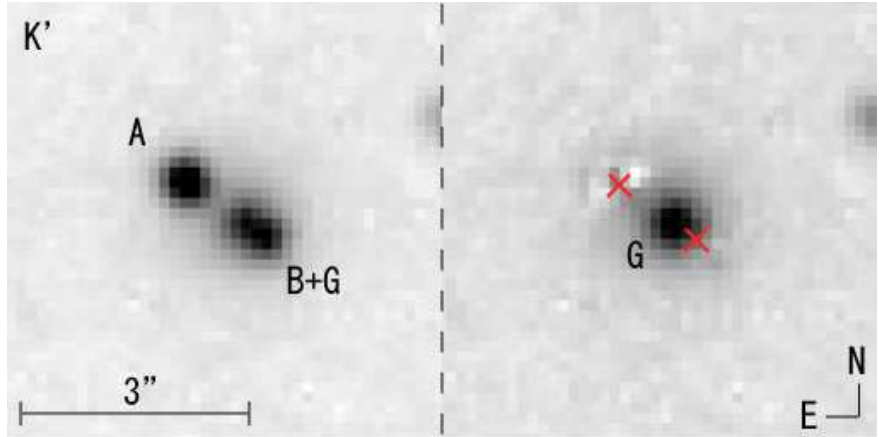}
\caption{The NIRC $K'$-band image of SDSS~J0806+2006.  The image scale
 is 0\farcs15 ${\rm pixel^{-1}}$. In the right panel, components A and
 B are subtracted. The lens galaxy can be clearly seen even before
 the PSF subtraction, as well as in the PSF-subtracted image. 
\label{fig:keck_0806}}
\end{figure}

We obtained $V$, $R$ and $I$-band images of SDSS~J0806+2006 using the 
8k mosaic CCD camera at the UH88 telescope, on 2004 December 16. 
The pixel scale of the 8k mosaic CCD camera is 0\farcs235
${\rm pixel^{-1}}$. The exposure time was 360 sec for each band. 
The typical seeing in the exposures was ${\sim}0\farcs7$, about a 
half of the typical seeing size of the SDSS (${\sim}1\farcs5$). 
Bias-subtracted and flat-field corrected images are shown in the upper 
panels of Figure~\ref{fig:uh88_0806}. We also obtained an $H$-band 
image using the QUick InfraRed Camera (QUIRC) of the UH88 telescope 
on 2005 February 21 and a $K'$-band image using the Near InfraRed Camera 
\citep[NIRC;][]{matthews94} of the Keck I telescope on 2005
April 24. The pixel scale of QUIRC is $0\farcs189$ ${\rm pixel^{-1}}$, 
and that of the NIRC is $0\farcs15$ ${\rm pixel^{-1}}$. The seeing was
${\sim}0\farcs7$ in the QUIRC observation, and it was ${\sim}0\farcs4$ in 
the NIRC observation. The total exposure time was 720 sec and 900 sec 
for $H$-band imaging and $K'$-band imaging, respectively. The $H$-band
image is shown in Figure~\ref{fig:uh88_0806}, and the $K'$-band image
is shown in Figure~\ref{fig:keck_0806}. 

All the images ($VRIHK'$) clearly show two stellar components; 
we name these two stellar components A (eastern component) and B 
(western component), with A being the brighter component.  
To determine if there are any extended objects in between the stellar
components, we subtracted point spread functions (PSFs) from the raw 
$VRIH$ images, adopting a nearby star as a template for the PSF. 
The results are shown in the lower panels of Figure~\ref{fig:uh88_0806};
there is clearly residual flux between components A and B in all four
images. This object (named G) is quite red, with colors of 
$V-R{\sim}1.1$ and $R-I{\sim}1.1$ that are similar to those of
an early-type galaxy at $z_l{\sim}0.6$ \citep{fukugita95}.  Such 
a redshift is consistent with the redshift of a \ion{Mg}{2} absorption 
line system in the spectra of the quasars (see \S \ref{sec:0806spc}) 
and an estimate based on the Faber-Jackson relation 
(see \S \ref{sec:0806model}). Therefore, we conclude that component G
is the lens galaxy. In the higher resolution NIRC $K'$-band image
(left panel of Figure~\ref{fig:keck_0806}), we can identify component
G between components A and B even before PSF subtraction. We fit this
image using GALFIT \citep{peng02} to find that G is well-modeled by a
de Vaucouleurs profile of ellipticity $e=0.18$ and major axis position
angle $\theta_e=51^\circ$.

The astrometry and photometry of components A, B and G are summarized in 
Table \ref{tab:0806}. The optical images were calibrated using the 
standard star PG 0231+051 \citep{landolt92} and the $H$-band image was 
calibrated using the standard star FS 21 \citep{hawarden01}. We lack a
photometric standard stars for the NIRC $K'$-band image, thus  
the data were used only for astrometry and the flux ratio constraints 
in the lens models.
When we fit the NIRC $K'$-band image using a model consisting of 
the two quasar images and the lens galaxy, we find a quasar flux 
ratio of 0.53 as compared to the mean flux ratio of 0.67 
(derived from fitting only the PSF models) for the optical 
bands that could be created by a modest amount of dust extinction or 
chromatic microlensing. We report the astrometry from using GALFIT to 
simultaneously fit components A, B and G in the NIRC $K'$-band image. 
The angular separation of components A and B is $1\farcs403\pm0\farcs012$.

%%%%%%%%%%%%%%%%%%%%%%%%%%%%%%%%%%%%%%%%%%%%%%%%
\subsection{Spectroscopic Observations}\label{sec:0806spc}
%%%%%%%%%%%%%%%%%%%%%%%%%%%%%%%%%%%%%%%%%%%%%%%%

\begin{figure}
\epsscale{.85}
\plotone{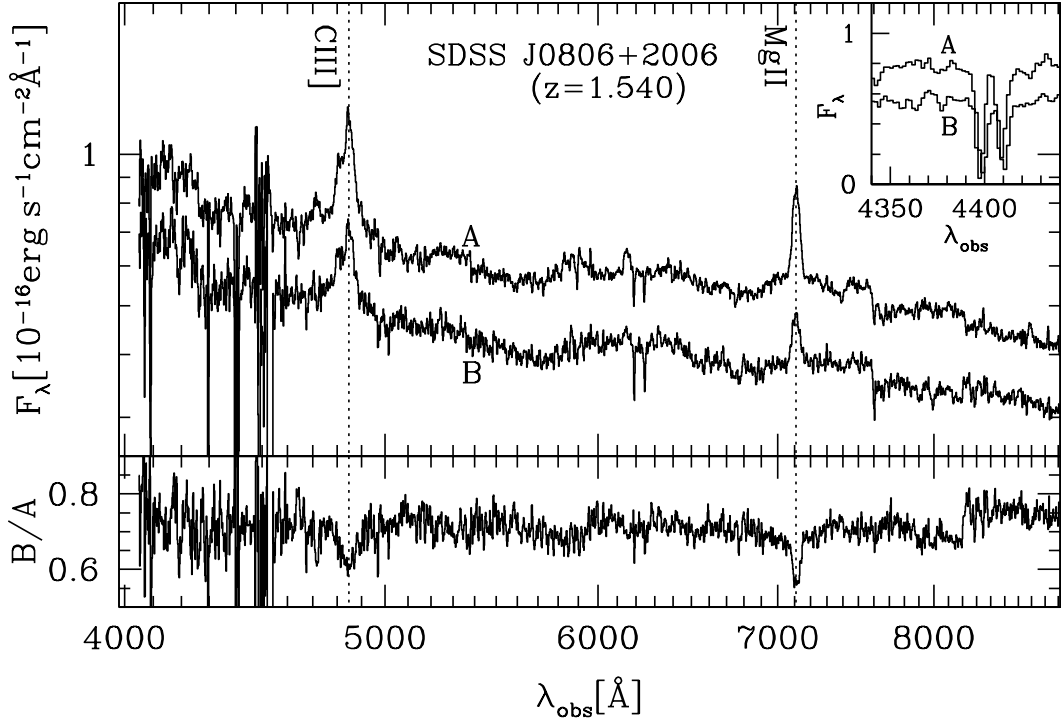}
\caption{The (binned) spectra of components A and B of SDSS~J0806+2006 taken
 with the ESI on the Keck II. The spectral resolution is R$\sim$27000.
 The vertical dotted lines represent the positions of 
 emission lines redshifted to $z = 1.534$ of \ion{C}{3]} (1908.73 {\AA}) and 
 \ion{Mg}{2} (2798.75 {\AA}), respectively. Both components clearly  
 have the identical redshift. The lower panel shows the spectral flux ratio; 
 it is almost constant and is consistent with the mean photometric
 flux ratio (0.67).  In the inset, we show the \ion{Mg}{2} absorption
 line system at $z=0.573$, which is probably associated with the lens
 galaxy. The data have bad columns around ${\sim}4500$ \AA.
\label{fig:spec0806}}
\end{figure}

A spectroscopic observation of SDSS~J0806+2006 was conducted with the
Keck II telescope on 2005 April 12 in ${\sim}1\farcs0$ seeing. We used
the echellette mode of the Echellette Spectrograph and Imager
\citep[ESI;][]{sutin97,sheinis02} with the MIT-LL 2048$\times$4096 CCD
camera. The spectral range was  3900{\,\AA} to 11,000{\,\AA} at a 
spectral resolution of $11.4~{\rm km\,s^{-1}pixel^{-1}}$ (R$\sim$27000). 
The exposure time was 900 sec. The $1\farcs0$-wide slit
was oriented to observe components A and B simultaneously. 
The spectra of each component was extracted using the standard method
(summing the fluxes in a window around the position of each component
and subtracting the sky using neighboring windows on either side of the trace). 
We show the binned spectra in Figure \ref{fig:spec0806}.
Both spectra have the \ion{C}{3]} and \ion{Mg}{2} emission lines at 
the same wavelength, with an estimated velocity difference of 
$20{\pm}70$~km s$^{-1}$ for the \ion{Mg}{2} emission line. 
We summarize the  redshifts calculated from
these emission lines in Table \ref{tab:0806em}. In addition, the
spectral energy distributions (SEDs) are also very similar. The ratio of
the spectra, which is shown in Figure \ref{fig:spec0806}, is almost
constant (${\sim}0.7$) for a wide range of wavelengths, and it is consistent
with the mean flux ratio of the optical images (0.67). Note, however,
that the emission line flux ratios appear to differ slightly from that
of the continuum, suggesting that there is some microlensing of the
continuum by the stars in the lens galaxy. 

In addition to the quasar emission lines, we found a strong 
\ion{Mg}{2} absorption line system at $z=0.573$ in both 
spectra (see the inset of Figure \ref{fig:spec0806}). We also 
found \ion{Ca}{2} H and K ($\sim6200$ \AA) and \ion{Fe}{2} 
absorption lines ($\sim4100$ \AA) at $z=0.573$ in the spectra. 
The fact that the redshift of this absorption line system is close 
to the estimated redshift (see \S \ref{sec:0806img}) of the probable 
lens galaxy (component G) and that \ion{Mg}{2} absorption lines
are frequently associated with galaxies  \citep[e.g.,][]{bergeron91}
suggests that the absorption is due to the lens galaxy. If this is the
case, the lens redshift should be $z_l{\simeq}0.573$; we adopt this
value for the mass models presented in the next subsection.

%%%%%%%%%%%%%%%%%%%%%%%%%%%%%%%%%%%%%%%%%%%%%%%%
\subsection{Mass Modeling}\label{sec:0806model}
%%%%%%%%%%%%%%%%%%%%%%%%%%%%%%%%%%%%%%%%%%%%%%%%

To explore the lensing hypothesis further, we modeled the system using
the two standard mass models: a Singular Isothermal Sphere with an 
external shear (SIS+shear) model, and a Singular Isothermal Ellipsoid (SIE) model. 
Both models have eight parameters: the Einstein radius $R_{\rm E}$, the
shear $\gamma$ or ellipticity $e$ and its position angle ($\theta_\gamma$ or $\theta_e$), 
the position of the lens galaxy, and the position and flux of the source
quasar.  We have only eight constraints (the positions of A, B and G,
and the fluxes of A and B), so we should be able to fit the data 
perfectly.  We should, however, be able to do so with sensible values
for the shear or ellipticity of the models.  
We used the component positions from Table~\ref{tab:0806} and the flux ratio 
of 0.53 derived from the GALFIT model of the NIRC $K'$-band image 
including the two stellar components and the lens galaxy.

We used {\it lensmodel} \citep{keeton01} to determine the model 
parameters, with the results summarized in Table~\ref{table:model0806}.  
The required ellipticities of $\gamma=0.03$ or $e=0.08$ are typical 
of other lensed
systems and roughly consistent with that of the lens galaxy ($e=0.18$),
 but there is a significant misalignment between the position
angle of the major axis of the models ($\theta_e{\simeq}-28^\circ$) 
and that of the lens galaxy ($\theta_e{=}51^\circ$). 
This might imply that the system is affected by a strong external 
perturbation \citep{keeton98}. Indeed, there are at least four galaxies 
within a 16$''$ radius of the lens (named G1--4, with the nearest 
galaxy only 4$''$ from the lens; see Table~\ref{tab:0806gal}), so
the position angle of the models may be a compromise between that
of the lens galaxy and the shear induced by G1. 
Finally, these models predict that the time delay between the images is
${\sim}32h^{-1}$days.

We can estimate the redshift of the lens galaxy based on the
Faber-Jackson relation \citep{faber76}. From the Table~3 
of \citet{rusin03}, we estimate that the lens magnitude should be 
$R{\sim}20.5$ for $\Delta{\theta}=1\farcs40$, $z_s{=}1.54$, and 
assuming $z_l=0.57$. 
While this is somewhat brighter than the observed $R$-band 
magnitude of galaxy G ($R=21.2$), it is roughly consistent given the
$\pm0.6$~mag scatter \citep{rusin03}. 
The rough agreement of this estimate further supports for a 
lens galaxy redshift of $z_l{\simeq}0.573$.

%%%%%%%%%%%%%%%%%%%%%%%%%%%%%%%%%%%%%%%%%%%%%%%%
%%%%%%%%%%%%%%%%%%%%%%%%%%%%%%%%%%%%%%%%%%%%%%%%
%%%%%%%%%%%%%%%%%%%%%%%%%%%%%%%%%%%%%%%%%%%%%%%%
\section{SDSS~J1353+1138}\label{sec:1353}
%%%%%%%%%%%%%%%%%%%%%%%%%%%%%%%%%%%%%%%%%%%%%%%%
%%%%%%%%%%%%%%%%%%%%%%%%%%%%%%%%%%%%%%%%%%%%%%%%
%%%%%%%%%%%%%%%%%%%%%%%%%%%%%%%%%%%%%%%%%%%%%%%%

%%%%%%%%%%%%%%%%%%%%%%%%%%%%%%%%%%%%%%%%%%%%%%%%
\subsection{Imaging Observations}\label{sec:1353img}
%%%%%%%%%%%%%%%%%%%%%%%%%%%%%%%%%%%%%%%%%%%%%%%%

\begin{figure}
\epsscale{.95}
\plotone{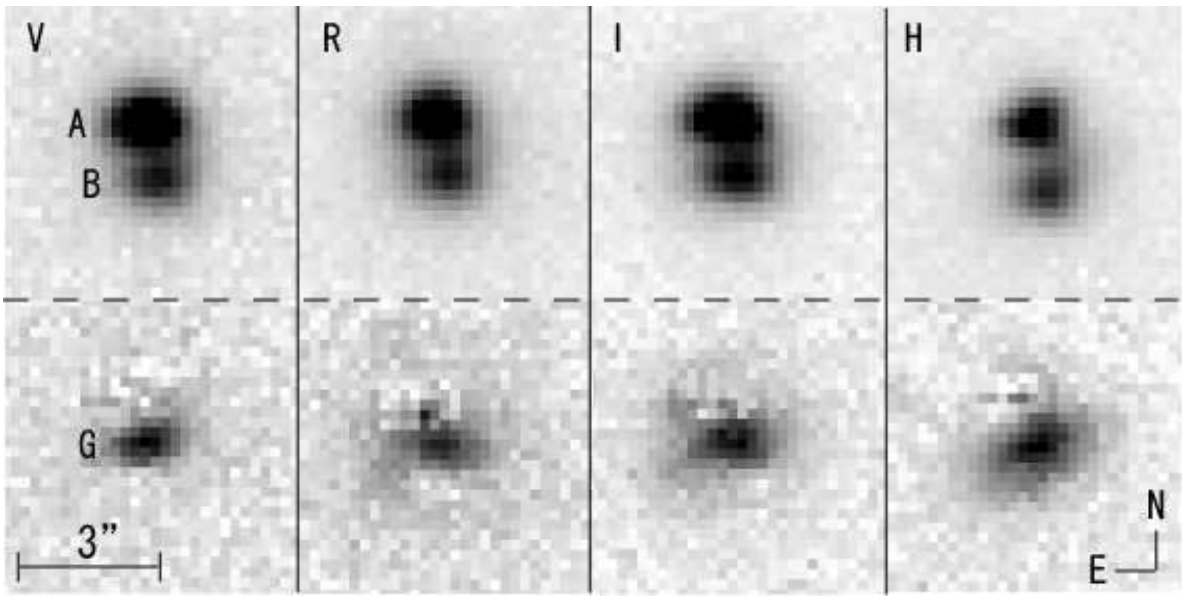}
\caption{Same as Figure \ref{fig:uh88_0806}, but for SDSS~J1353+1138. 
 The lens galaxy is also clearly seen in the PSF-subtracted
 images. 
\label{fig:uh88_1353}}
\end{figure}

We obtained $V$, $R$, and $I$-band images of SDSS~J1353+1138 
using the 8K mosaic CCD camera of the UH88 telescope on 2004
May 25 in ${\sim}1\farcs0$ seeing.  The exposure times were
120 sec for $V$ and 180 sec for $R$ and $I$, respectively. 
We also acquired  an $H$-band image on the UH88 telescope
using QUIRC on 2005 February 21. The exposure time was 720 sec, and the 
seeing was ${\sim}0\farcs7$. The images, shown in the upper panels 
of Figure~\ref{fig:uh88_1353}, clearly show two stellar components, 
which we named A (northern and brighter component) and B (southern component). 
Although it is not obvious in the raw $H$-band image of Figure
\ref{fig:uh88_1353}, an extended object can be seen between components A and
B even before the PSF subtraction.  We also obtained $g$ and $i$-band
images using the Magellan Instant Camera (MagIC) at the LC6.5m telescope on 2005 April 15. 
The pixel scale of MagIC is
$0\farcs069$ ${\rm pixel^{-1}}$. The seeing was ${\sim}0\farcs6$, and the 
exposure time was 360 sec for $g$-band and 480 sec for $i$-band.  

\begin{figure}
\epsscale{.6}
\plotone{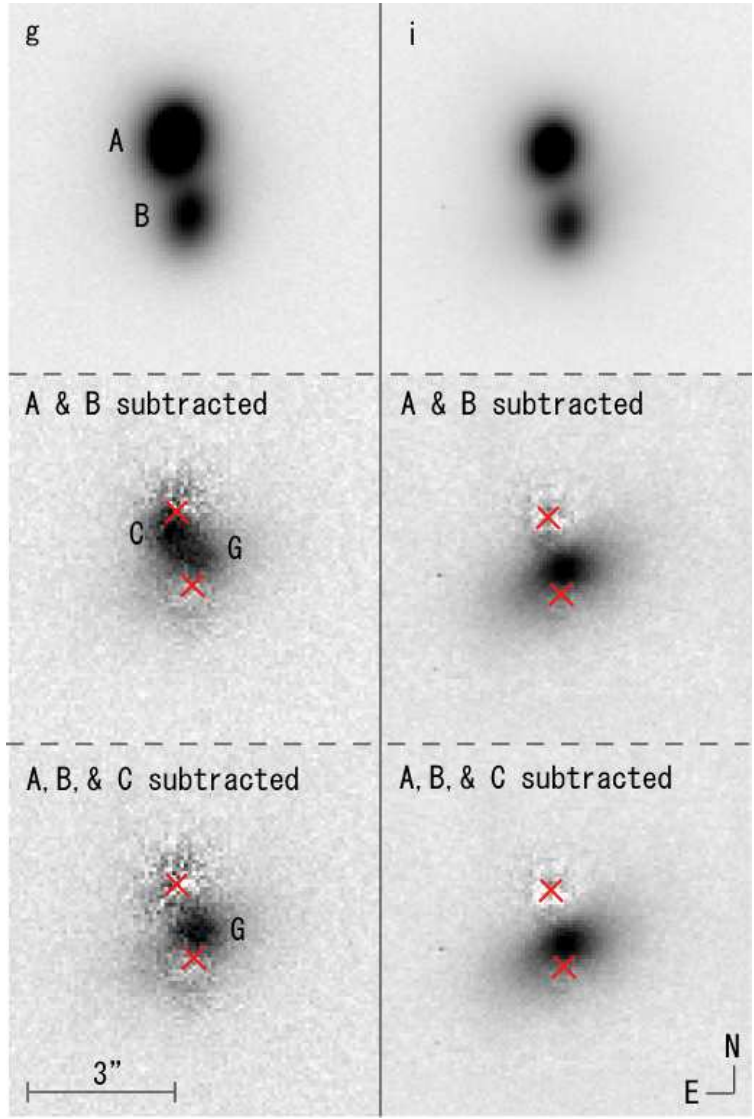}
\caption{The MagIC $gi$-band images of SDSS~J1353+1138. 
 The image scale is 0\farcs069 ${\rm pixel^{-1}}$. The top panels
 show the original images, the middle panels show the residuals after 
 subtracting components A and B, and the bottom panels show the 
 residuals after subtracting components A, B and C.  Note the
 relatively blue color of component C compared to the quasar 
 images and the lens galaxy.  
\label{fig:mag_1353}}
\end{figure}

After subtracting PSF models for the two stellar components, we 
clearly detect an extended red object positioned between them in all
the residual images.  These residual images are shown in 
Figure \ref{fig:uh88_1353} and \ref{fig:mag_1353} for the UH88
and Magellan data, respectively.  We identify this extended
object, which we denote as G, with the lens galaxy.  We 
fit the $i$-band image using GALFIT to find that the lens galaxy
is well-modeled with a de Vaucouleurs profile of ellipticity
$e=0.50$ and major axis position angle $\theta_e{=}-62^\circ$.
In addition to component G, there is an additional object
which we named component C near image A.  It is most
easily seen in the $g$-band image after subtracting 
components A and B (see the middle panel of Figure 
\ref{fig:mag_1353}), but it can also be seen at the same
position in the $i$-band image.

There are four possible interpretations of component C. First, it
could be a chance superposition of a foreground star. It is 
well fit by the PSF (see the bottom panels of Figure \ref{fig:mag_1353}), 
and our lens models require only components A, B and G to reproduce
the system with reasonable parameter values (see \ref{sec:1353model}).  
Second, it could be an object related to the lens galaxy. 
However, lens models with significant extra mass at the position of component
C generally fit the data very badly.  Third, it could be a third image of 
the quasar.  This seems unlikely since its colors are very different from
components A and B (see Table \ref{tab:1353}). Moreover, identifying
it as a quasar image makes little sense as a lens geometry since it is
in the wrong place to be a central, odd image (which should lie between
components B and G) or to be part of a four-image system in which one
of the A/B/C components is an unresolved image pair.  Fourth, it could be 
emission from the host galaxy of the source quasar. Probably only the first 
possibility is permitted, although we cannot 
completely exclude the other possibilities.  

We summarize astrometry and photometry of components A, B, C and G 
in Table \ref{tab:1353}. The optical images were calibrated using 
the standard stars PG 1528+062 \citep{landolt92} and SA 107-351 \citep{smith02}, 
and the $H$-band image was calibrated using the standard star 
FS 21 \citep{hawarden01}. 
The flux ratio of components A and B changes significantly with wavelength,
from a mean flux ratio in the optical of $0.35$ to an $H$-band flux 
ratio of 0.54.  The angular
separation of components A and B is $1\farcs406\pm0\farcs007$. 

%%%%%%%%%%%%%%%%%%%%%%%%%%%%%%%%%%%%%%%%%%%%%%%%
\subsection{Spectroscopic Observations}\label{sec:1353spc}
%%%%%%%%%%%%%%%%%%%%%%%%%%%%%%%%%%%%%%%%%%%%%%%%

\begin{figure}
\epsscale{.85}
\plotone{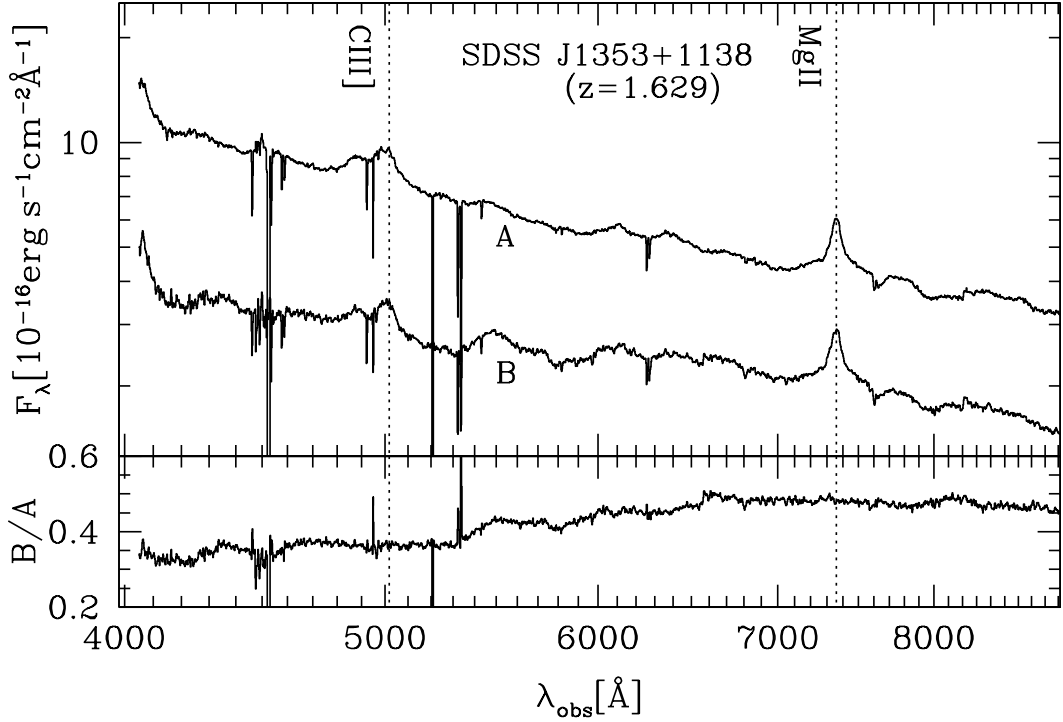}
\caption{Same as Figure \ref{fig:spec0806}, but for the (binned) spectra of
  SDSS~J1353+1138 are shown. The spectral resolution is R$\sim$27000.
  We find that both components
  have the identical redshift. The spectral flux ratio is also almost
  constant and is consistent with the mean photometric flux ratio
  (0.35). There are three \ion{Mg}{2} absorption line systems at
  $z=1.238$ (${\sim}6260$ {\AA}), $z=0.904$ (${\sim}5330$ {\AA}), and
  $z=0.637$ (${\sim}4580$ {\AA}) with their \ion{Fe}{2} absorption lines, 
  but they probably arise from the intervening galaxies, not the lens galaxy.  
  The data have bad columns around ${\sim}4500$ \AA. 
\label{fig:spec1353}}
\end{figure}

We used ESI on the Keck II telescope to obtain spectra of the two components 
of SDSS~J1353+1138 on 2005 April 12.  The instrumental set up was the
same as in \S\ref{sec:0806spc} and we used an exposure time of 600~sec.
The binned spectra are shown in Figure \ref{fig:spec1353}. Both 
components have \ion{C}{3]} and \ion{Mg}{2} emission lines at 
identical wavelengths, with a velocity difference of only 
$10{\pm}50$~km s$^{-1}$ for the \ion{Mg}{2} emission line. 
We summarize the redshifts derived from the emission lines 
in Table \ref{tab:1353em}. 

The flux ratio of the spectra is nearly constant (${\sim}0.4$) 
and consistent with that in the optical images ($\sim 0.35$).
There is, however, a slight increase in the ratio redwards of
about 5300 {\AA} that may be due to contamination of the 
spectrum of component B by emission from the lens galaxy. 
Assuming this feature is due to the 4000{\AA} break of the
lens galaxy, it implies a lens redshift of $z_l \sim 0.3$.
Since the colors of the lens galaxy of $V-R{\sim}0.3$, $R-I{\sim}1.0$, 
and $g-i{\sim}1.1$ are roughly consistent with those of an early-type
galaxy at $z=0.1{\sim}0.3$ \citep{fukugita95}, we adopt $z_l=0.3$ for
mass modeling. 

The quasar spectra also contain strong \ion{Mg}{2} absorption line systems 
at $z=1.238$, $0.904$, and $0.637$, with their \ion{Fe}{2} absorption lines.  
These are unlikely to be associated
with the lens galaxy, given the large difference from the lens
redshift estimated from the colors, and also given the fact that many 
(unlensed) quasars show \ion{Mg}{2} absorption. We note that 
an \ion{Mg}{2} absorption line due to a $z \simeq 0.3$ lens is
undetectable in our spectra because it would lie beyond the
atmospheric cutoff. 

%%%%%%%%%%%%%%%%%%%%%%%%%%%%%%%%%%%%%%%%%%%%%%%%
\subsection{Mass Modeling}\label{sec:1353model}
%%%%%%%%%%%%%%%%%%%%%%%%%%%%%%%%%%%%%%%%%%%%%%%%

We modeled SDSS~J1353+1138 in the same way as in \S \ref{sec:0806model}.
The positions of the quasars and lens galaxy are taken from Table
\ref{table:model1353} (neglecting component C). For
the flux ratio of components A and B, we adopt $0.31$, which was 
derived from the GALFIT model of the MagIC $i$-band image.
The results are summarized in Table \ref{table:model1353}. 
The derived shear and ellipticity of $\gamma=0.05$ and $e=0.15$, 
are again typical for a lensed quasar system.
The derived ellipticity (0.15) appears to be significantly smaller than
the ellipticity of the observed light profile ($e{=}0.50$), 
but this difference is commonly seen in the lensed quasar systems
\citep{keeton98}. However, in addition to the disagreement of 
the ellipticity, there is a significant misalignment between the position
angle of the models ($\theta_e{\simeq}-34^\circ$) and that of the
lens galaxy ($\theta_e{=}-62^\circ$).  While this might be
due to external perturbations, there are no nearby galaxies 
that can easily explain the misalignment.  It could be due
to component C, but we find that simple tests of lens models with mass
located at the position of component C generally fail to fit the
data. The time delay between A and B is predicted to be
${\sim}16h^{-1}$days, assuming the lens redshift of $z_l=0.3$.  

As in \S\ref{sec:0806model} we also estimated the lens redshift based on 
the Faber-Jackson relation. Again we used Table~3 of \citet{rusin03},  
$\Delta{\theta}=1\farcs41$, and $z_s{=}1.63$ to estimate that the lens 
magnitude should be $R{\sim}19.2$, assuming ${z_l}=0.3$. 
This estimate is in rough agreement with the observed 
$R$-band magnitude of galaxy G ($R=18.8$), again supporting for
a lens redshift of $z_l\sim0.3$.

%%%%%%%%%%%%%%%%%%%%%%%%%%%%%%%%%%%%%%%%%%%%%%%%
%%%%%%%%%%%%%%%%%%%%%%%%%%%%%%%%%%%%%%%%%%%%%%%%
%%%%%%%%%%%%%%%%%%%%%%%%%%%%%%%%%%%%%%%%%%%%%%%%
\section{Summary}\label{sec:conc}
%%%%%%%%%%%%%%%%%%%%%%%%%%%%%%%%%%%%%%%%%%%%%%%%
%%%%%%%%%%%%%%%%%%%%%%%%%%%%%%%%%%%%%%%%%%%%%%%%
%%%%%%%%%%%%%%%%%%%%%%%%%%%%%%%%%%%%%%%%%%%%%%%%

We report the discovery of two doubly-imaged quasar lenses, 
SDSS~J0806+2006 and SDSS~1353+1138.  Both were selected from the SDSS 
spectroscopic quasar sample as lensed quasar candidates and confirmed 
in subsequent imaging and spectroscopic observations.  
SDSS~J0806+2006 consists of two $z_s=1.540$ quasar images separated by 
$\Delta{\theta}=1\farcs40$ lensed by a galaxy at $z_l\simeq0.6$.  
The lens galaxy is closer to the fainter image as expected, and its
redshift, as suggested by its magnitude, colors, and the presence of a
\ion{Mg}{2} absorption feature, is $z_l=0.573$. Several nearby galaxies 
may perturb this system and indicate that the lens galaxy is part of a 
small group. SDSS~1353+1138 consists of two $z_s=1.629$ quasar images
separated  by $\Delta{\theta}=1\farcs41$ with a lens galaxy at
$z_l{\sim}0.3$. The redshift of the lens galaxy is estimated based
on its magnitude, colors, and the spectral  flux ratio between the two
quasar images. There is an additional component, which we have labeled
C, superposed on this system whose nature is presently unexplained.
Observations using the {\em Hubble Space Telescope} are probably
required to clarify its role in the lensed quasar system. 

%In summary, we have identified the two new gravitational lens candidates,
%and presented convincing evidence for the lensing hypotheses. These new
%lenses will serve as an important element of the statistical sample of
%lensed quasars in the SDSS, since they were selected via the
%homogeneous selection. 

\acknowledgments

N.~I. and M.~O. are supported by JSPS through JSPS Research Fellowship
for Young Scientists. A portion of this work was also performed under
the auspices of the U.S. Department of Energy,
National Nuclear Security Administration by the University of California,
Lawrence Livermore National Laboratory under contract No. W-7405-Eng-48. 

Some of the data presented herein were obtained at the W.M. Keck Observatory, 
which is operated as a scientific partnership among the California Institute 
of Technology, the University of California and the National Aeronautics and 
Space Administration. The Observatory was made possible by the generous 
financial support of the W.M. Keck Foundation. 

Funding for the creation and distribution of the SDSS Archive has been 
provided by the Alfred P. Sloan Foundation, the Participating Institutions, 
the National Aeronautics and Space Administration, the National Science Foundation, 
the U.S. Department of Energy, the Japanese Monbukagakusho, and the Max Planck 
Society. The SDSS Web site is http://www.sdss.org/. 

The SDSS is managed by the Astrophysical Research Consortium (ARC) for the 
Participating Institutions. The Participating Institutions are The University 
of Chicago, Fermilab, the Institute for Advanced Study, the Japan Participation 
Group, The Johns Hopkins University, the Korean Scientist Group, Los Alamos 
National Laboratory, the Max-Planck-Institute for Astronomy (MPIA), the 
Max-Planck-Institute for Astrophysics (MPA), New Mexico State University, 
University of Pittsburgh, University of Portsmouth, Princeton University, 
the United States Naval Observatory, and the University of Washington.

\clearpage

%%%%%%%%%%%%%%%%%%%%%%%%%%%%%%%%%%%%%%%%%%%%%%%%%%%%%%%%%%%%%%%%%%%%%%%
\begin{deluxetable}{crrrrrr}
\rotate
\tablewidth{0pt}
\tablecaption{ASTROMETRY AND PHOTOMETRY OF SDSS~J0806+2006\label{tab:0806}}
\tablehead{\colhead{Object} & \colhead{$\Delta$R.A.(${}''$)\tablenotemark{a}} &
 \colhead{$\Delta$Dec.(${}''$)\tablenotemark{a}} &  
 \colhead{$V$\tablenotemark{b}} & \colhead{$R$\tablenotemark{b}} &
 \colhead{$I$\tablenotemark{b}} & \colhead{$H$\tablenotemark{b}}} 
\startdata
A &  0.000{$\pm$}0.005 &  0.000{$\pm$}0.005 & 19.23{$\pm$}0.01 & 18.93{$\pm$}0.01 & 18.54{$\pm$}0.01 & 16.87{$\pm$}0.01 \\
B & $-$1.136{$\pm$}0.007 & $-$0.823{$\pm$}0.007 & 19.82{$\pm$}0.02 & 19.36{$\pm$}0.02 & 18.84{$\pm$}0.01 & 17.34{$\pm$}0.02 \\
G & $-$0.811{$\pm$}0.015 & $-$0.574{$\pm$}0.015 & 22.27{$\pm$}0.07 & 21.20{$\pm$}0.04 & 20.16{$\pm$}0.03 & 17.90{$\pm$}0.05 \\
\enddata
%\tablecomments{}
\tablenotetext{a}{Measured in the Keck NIRC $K'$-band image using GALFIT. The celestial coordinates 
of component A are R.A.$=121\fdg59875$ and Decl.$=20\fdg10886$ (J2000).}
\tablenotetext{b}{PSF magnitudes for the stellar objects and a $2\farcs0$ radius aperture magnitude for the extended object.
  The errors do not include the photometric uncertainty of the standard star or the uncertainty due to the PSF subtraction.}
\end{deluxetable}
%%%%%%%%%%%%%%%%%%%%%%%%%%%%%%%%%%%%%%%%%%%%%%%%%%%%%%%%%%%%%%%%%%%%%%%

\clearpage

%%%%%%%%%%%%%%%%%%%%%%%%%%%%%%%%%%%%%%%%%%%%%%%%%%%%%%%%%%%%%%%%%%%%%%%
\begin{deluxetable}{lccccccc} 
\rotate
\tablecolumns{8} 
\tablewidth{0pc} 
\tablecaption{EMISSION LINES OF SDSS~J0806+2006 SPECTRA\label{tab:0806em}} 
\tablehead{ 
\colhead{} & \multicolumn{3}{c}{Component A} & \colhead{} & 
\multicolumn{3}{c}{Component B} \\ 
\cline{2-4} \cline{6-8} \\ 
\colhead{Line({\,\AA})} & \colhead{${\lambda}_{obs}$({\,\AA})} & \colhead{FWHM({\,\AA})} & \colhead{Redshift} & \colhead{} &
\colhead{${\lambda}_{obs}$({\,\AA})} & \colhead{FWHM({\,\AA})} & \colhead{Redshift} }
\startdata 
\ion{C}{3]}(1908.73) & 4847.18 & 42.4 & 1.5395$\pm$0.0005 & & 4847.99  & 43.8 & 1.5399$\pm$0.0007 \\
\ion{Mg}{2}(2798.75) & 7110.25 & 48.2 & 1.5405$\pm$0.0004 & & 7109.44  & 52.0 & 1.5402$\pm$0.0006 \\
\enddata 
\end{deluxetable} 
%%%%%%%%%%%%%%%%%%%%%%%%%%%%%%%%%%%%%%%%%%%%%%%%%%%%%%%%%%%%%%%%%%%%%%%

\clearpage

%%%%%%%%%%%%%%%%%%%%%%%%%%%%%%%%%%%%%%%%%%%%%%%%%%%%%%%%%%%%%%%%%%%%%%%
\begin{deluxetable}{cccccc}
\tablewidth{0pt}
\tablecaption{LENS MODELING: SDSS~J0806+2006\label{table:model0806}}
\tablehead{\colhead{Model} & \colhead{$R_{\rm E}$(${}''$)} &
 \colhead{$\gamma$ or $e$} &
 \colhead{$\theta_\gamma$ or $\theta_e$\tablenotemark{a}} & \colhead{$\Delta
 t$[$h^{-1}$day]\tablenotemark{b}} & \colhead{{$\mu_{\rm tot}$} \tablenotemark{c}}} 
\startdata
SIS+shear & $0.68$ & $0.03$ & $-28.0$ & $31.6$ & 4.63 \\
SIE         & $0.69$ & $0.08$ & $-27.9$ & $32.5$ & 4.62 \\
\enddata
%\tablecomments{}
\tablenotetext{a}{Each position angle is measured East of North.}
\tablenotetext{b}{The lens redshift is assumed to be $z=0.57$.}
\tablenotetext{c}{Predicted total magnification.}
\end{deluxetable}
%%%%%%%%%%%%%%%%%%%%%%%%%%%%%%%%%%%%%%%%%%%%%%%%%%%%%%%%%%%%%%%%%%%%%%%

\clearpage

%%%%%%%%%%%%%%%%%%%%%%%%%%%%%%%%%%%%%%%%%%%%%%%%%%%%%%%%%%%%%%%%%%%%%%%
\begin{deluxetable}{crrr}
\tablewidth{0pt}
\tablecaption{POSITIONS OF GALAXIES AROUND SDSS~J0806+2006\label{tab:0806gal}}
\tablehead{\colhead{Object} & \colhead{$\Delta$R.A.(${}''$)\tablenotemark{a}} &
 \colhead{$\Delta$Dec.(${}''$)\tablenotemark{a}} & \colhead{{$\ast$}/G\tablenotemark{b}}} 
\startdata
G1 &  $-$3.766 &    0.965 & 0.11 \\
G2 &     5.957 &    2.451 & 0.03 \\
G3 &     3.480 & $-$8.784 & 0.09 \\
G4 & $-$13.770 & $-$7.970 & 0.07 \\
\enddata
%\tablecomments{}
\tablenotetext{a}{Positions relative to component A, measured in the Keck NIRC $K'$-band image}
\tablenotetext{b}{$K'$-band flux ratio between each object and component G }
\end{deluxetable}
%%%%%%%%%%%%%%%%%%%%%%%%%%%%%%%%%%%%%%%%%%%%%%%%%%%%%%%%%%%%%%%%%%%%%%%

\clearpage

%%%%%%%%%%%%%%%%%%%%%%%%%%%%%%%%%%%%%%%%%%%%%%%%%%%%%%%%%%%%%%%%%%%%%%%
\begin{deluxetable}{crrrrrrrr}
\rotate
\tablewidth{0pt}
\tablecaption{ASTROMETRY AND PHOTOMETRY OF SDSS~J1353+1138\label{tab:1353}}
\tablehead{\colhead{Object} & \colhead{$\Delta$R.A.(${}''$)\tablenotemark{a}} &
 \colhead{$\Delta$Dec.(${}''$)\tablenotemark{a}} & 
 \colhead{$V$\tablenotemark{b}} & \colhead{$R$\tablenotemark{b}} & \colhead{$I$\tablenotemark{b}}  & 
 \colhead{$H$\tablenotemark{b}} & \colhead{$g$\tablenotemark{b}} & 
 \colhead{$i$\tablenotemark{b}} }
\startdata
A &  0.000{$\pm$}0.003 &  0.000{$\pm$}0.004 & 17.08{$\pm$}0.01 & 16.74{$\pm$}0.01 & 16.44{$\pm$}0.01 & 15.16{$\pm$}0.01 & 16.91{$\pm$}0.01 & 16.77{$\pm$}0.01 \\
B & $-$0.267{$\pm$}0.003 & $-$1.380{$\pm$}0.004 & 18.63{$\pm$}0.02 & 17.63{$\pm$}0.01 & 17.43{$\pm$}0.01 & 15.83{$\pm$}0.02 & 18.50{$\pm$}0.01 & 17.88{$\pm$}0.01 \\
C &  0.107{$\pm$}0.044 & $-$0.358{$\pm$}0.044 & \nodata\phn      & \nodata\phn      & \nodata\phn      & \nodata\phn      & 20.96{$\pm$}0.06 & 21.67{$\pm$}0.10 \\
G & $-$0.255{$\pm$}0.008 & $-$1.041{$\pm$}0.008 & 19.06{$\pm$}0.03 & 18.80{$\pm$}0.02 & 17.80{$\pm$}0.01 & 16.16{$\pm$}0.02 & 19.84{$\pm$}0.01 & 18.77{$\pm$}0.01 \\
\enddata
%\tablecomments{}
\tablenotetext{a}{Measured in the MagIC $i$-band image using GALFIT (except component C). The celestial coordinates of 
component A are R.A.$=208\fdg27646$ and Decl.$=11\fdg63467$ (J2000).}
\tablenotetext{b}{PSF magnitudes for the stellar objects and a $2\farcs0$ radius aperture magnitude for the extended object.
  The errors do not include the photometric uncertainty of the standard star or the uncertainty of the PSF subtraction.}
\end{deluxetable}
%%%%%%%%%%%%%%%%%%%%%%%%%%%%%%%%%%%%%%%%%%%%%%%%%%%%%%%%%%%%%%%%%%%%%%%

\clearpage

%%%%%%%%%%%%%%%%%%%%%%%%%%%%%%%%%%%%%%%%%%%%%%%%%%%%%%%%%%%%%%%%%%%%%%%
\begin{deluxetable}{lccccccc} 
\rotate
\tablecolumns{8} 
\tablewidth{0pc} 
\tablecaption{EMISSION LINES OF SDSS~J1353+1138 SPECTRA\label{tab:1353em}} 
\tablehead{ 
\colhead{} & \multicolumn{3}{c}{Component A} & \colhead{} & 
\multicolumn{3}{c}{Component B} \\ 
\cline{2-4} \cline{6-8} \\ 
\colhead{Line({\,\AA})} & \colhead{${\lambda}_{obs}$({\,\AA})} & \colhead{FWHM({\,\AA})} & \colhead{Redshift} & \colhead{} &
\colhead{${\lambda}_{obs}$({\,\AA})} & \colhead{FWHM({\,\AA})} & \colhead{Redshift} }
\startdata 
\ion{C}{3]}(1908.73) & 5007.61 & 75.6 & 1.6235$\pm$0.0006 & & 5008.08  & 75.6 & 1.6238$\pm$0.0006 \\
\ion{Mg}{2}(2798.75) & 7359.33 & 74.3 & 1.6295$\pm$0.0004 & & 7358.66  & 73.4 & 1.6293$\pm$0.0004 \\
\enddata 
\end{deluxetable} 
%%%%%%%%%%%%%%%%%%%%%%%%%%%%%%%%%%%%%%%%%%%%%%%%%%%%%%%%%%%%%%%%%%%%%%%

\clearpage

%%%%%%%%%%%%%%%%%%%%%%%%%%%%%%%%%%%%%%%%%%%%%%%%%%%%%%%%%%%%%%%%%%%%%%%
\begin{deluxetable}{cccccc}
\tablewidth{0pt}
\tablecaption{LENS MODELING: SDSS~J1353+1138\label{table:model1353}}
\tablehead{\colhead{Model} & \colhead{$R_{\rm E}$(${}''$)} &
 \colhead{$\gamma$ or $e$} &
 \colhead{$\theta_\gamma$ or $\theta_e$\tablenotemark{a}} & \colhead{$\Delta
 t$[$h^{-1}$day]\tablenotemark{b}} & \colhead{{$\mu_{\rm tot}$} \tablenotemark{c}}} 
\startdata
SIS+shear & $0.71$ & $0.05$ & $-35.1$ & $16.2$ & 3.81 \\
SIE         & $0.70$ & $0.15$ & $-34.3$ & $16.4$ & 3.75 \\
\enddata
%\tablecomments{}
\tablenotetext{a}{Each position angle is measured East of North.}
\tablenotetext{b}{The lens redshift is assumed to be $z=0.3$.}
\tablenotetext{c}{Predicted total magnification.}
\end{deluxetable}
%%%%%%%%%%%%%%%%%%%%%%%%%%%%%%%%%%%%%%%%%%%%%%%%%%%%%%%%%%%%%%%%%%%%%%%

\clearpage

%%%%%%%%%%%%%%%%%%%%%%%%%%%%%%%%%%%%%%%%%%%%%%%%%%%%%%%%%%%%%%%%%%%%%%%
%%%%%%%%%%%%%%%%%%%%%%%%%%%%%%%%%%%%%%%%%%%%%%%%%%%%%%%%%%%%%%%%%%%%%%%

\clearpage

%%%%%%%%%%%%%%%%%%%%%%%%%%%%%%%%%%%%%%%%%%%%%%%%%%%%%%%%%%%%%%%%%%%%%%%
%%%%%%%%%%%%%%%%%%%%%%%%%%%%%%%%%%%%%%%%%%%%%%%%%%%%%%%%%%%%%%%%%%%%%%%

\clearpage

%%%%%%%%%%%%%%%%%%%%%%%%%%%%%%%%%%%%%%%%%%%%%%%%%%%%%%%%%%%%%%%%%%%%%%%
%%%%%%%%%%%%%%%%%%%%%%%%%%%%%%%%%%%%%%%%%%%%%%%%%%%%%%%%%%%%%%%%%%%%%%%

\clearpage

%%%%%%%%%%%%%%%%%%%%%%%%%%%%%%%%%%%%%%%%%%%%%%%%%%%%%%%%%%%%%%%%%%%%%%%
%%%%%%%%%%%%%%%%%%%%%%%%%%%%%%%%%%%%%%%%%%%%%%%%%%%%%%%%%%%%%%%%%%%%%%%

\clearpage

%%%%%%%%%%%%%%%%%%%%%%%%%%%%%%%%%%%%%%%%%%%%%%%%%%%%%%%%%%%%%%%%%%%%%%%
%%%%%%%%%%%%%%%%%%%%%%%%%%%%%%%%%%%%%%%%%%%%%%%%%%%%%%%%%%%%%%%%%%%%%%%

\clearpage

%%%%%%%%%%%%%%%%%%%%%%%%%%%%%%%%%%%%%%%%%%%%%%%%%%%%%%%%%%%%%%%%%%%%%%%
%%%%%%%%%%%%%%%%%%%%%%%%%%%%%%%%%%%%%%%%%%%%%%%%%%%%%%%%%%%%%%%%%%%%%%%

\clearpage

%%%%%%%%%%%%%%%%%%%%%%%%%%%%%%%%%%%%%%%%%%%%%%%%%%%%%%%%%%%%%%%%%%%%%%%
%%%%%%%%%%%%%%%%%%%%%%%%%%%%%%%%%%%%%%%%%%%%%%%%%%%%%%%%%%%%%%%%%%%%%%%


\begin{thebibliography}{}

\bibitem[Abazajian et al.(2003)]{abazajian03}
Abazajian, K., et al. 2003, \aj, 126, 2081

\bibitem[Abazajian et al.(2004)]{abazajian04}
Abazajian, K., et al. 2004, \aj, 128, 502

\bibitem[Abazajian et al.(2005)]{abazajian05}
Abazajian, K., et al. 2005, \aj, 129, 1755

\bibitem[Adelman-McCarthy(2005)]{adelman05}
Adelman-McCarthy, J.~K., et al. 2005, 
\apjs, in press (astro-ph/0507711) 
 
\bibitem[Bade et al.(2003)]{bade97} 
Bade, N., Siebert, J., Lopez, S., Voges, W., \& 
Reimers, D. 1997, A\&A, 317, L13

\bibitem[Bergeron \& Boisse(1991)]{bergeron91}
Bergeron, J., \& Boisse, P. 1991, A\&A, 243, 344

\bibitem[Blanton et al.(2003)]{blanton03} 
Blanton, M. R., Lin, H., Lupton, R. H., Maley, F. M., Young, N., 
Zehavi, I., \& Loveday, J. 2003, \aj, 125, 2276

\bibitem[Browne et al.(2003)]{browne03} 
Browne, I. W. A., et al. 2003, \mnras, 341, 13

\bibitem[Burles et al.(2005)]{burles05}
Burles, S., et al. 2005, \aj, in preparation

\bibitem[Chae et al.(2002)]{chae02}
Chae, K.-H., et al.\ 2002, Phys. Rev. Lett., 89, 151301 
 
\bibitem[Eisenstein et al.(2001)]{eisenstein01} 
Eisenstein, D. J., et al. 2001, \aj, 122, 2267

\bibitem[Faber \& Jackson(1976)]{faber76} 
Faber, S. M. \& Jackson, R. E. 1976, \apj, 204, 668

\bibitem[Fukugita et al.(1990)]{fukugita90}
Fukugita, M., Futamase, T., \& Kasai, M.\ 1990, \mnras, 246, 24P

\bibitem[Fukugita et al.(1995)]{fukugita95} 
Fukugita, M., Shimasaku, K., \& Ichikawa, T. 1995, PASP, 107, 945

\bibitem[Fukugita et al.(1996)]{fukugita96}
Fukugita, M., Ichikawa, T., Gunn, J. E., Doi, M., 
Shimasaku, K., \& Schneider, D. P. 1996, \aj, 111, 1748

\bibitem[Gunn et al.(1998)]{gunn98} 
Gunn, J. E., et al. 1998, \aj, 116, 3040

\bibitem[Gunn et al.(2005)]{gunn05} 
Gunn, J. E., et al. 2005, \aj, submitted

\bibitem[Hawarden et al.(2001)]{hawarden01}
Hawarden, T.~G., Leggett, S.~K., Letawsky, M.~B., Ballantyne, D.~R., 
\& Casali, M.~M.\ 2001, \mnras, 325, 563 

\bibitem[Hogg et al.(2001)]{hogg01}
Hogg, D. W., Finkbeiner, D. P., Schlegel, D. J., 
\& Gunn, J. E.\ 2001, \aj, 122, 2129

\bibitem[Inada et al.(2003a)]{inada03a}
Inada, N., et al. 2003a, \aj, 126, 666

\bibitem[Inada et al.(2003b)]{inada03b}
Inada, N., et al. 2003b, \aj, submitted

\bibitem[Inada et al.(2003c)]{inada03c}
Inada, N., et al. 2003c, \nat, 426, 810

\bibitem[Inada et al.(2005)]{inada05}
Inada, N., et al. 2005, \aj, 130, 1967

\bibitem[Ivezi\'{c} et al.(2004)]{ivezic04}
Ivezi\'{c}, \v{Z}., et al. 2004, AN, 325, 583

\bibitem[Johnston et al.(2003)]{johnston03}
Johnston, D. E., et al. 2003, \aj, 126, 2281

\bibitem[Keeton(2001b)]{keeton01}
Keeton, C.~R.\ 2001b, preprint (astro-ph/0102340)

\bibitem[Keeton et al.(1998)]{keeton98}
Keeton, C.~R., Kochanek, C.~S., \& Falco, E.~E.\ 1998, 
\apj, 509, 561

\bibitem[Kochanek(1991)]{kochanek91} 
Kochanek, C.~S. 1991, \apj, 373, 354

\bibitem[Kochanek et al.(2004)]{kochanek04}
Kochanek, C. S., Schneider, P., Wambsganss, J., 2004, 
Part 2 of Gravitational Lensing: Strong, Weak \& Micro, 
Proceedings of the 33rd Saas-Fee Advanced Course, 
G. Meylan, P. Jetzer \& P. North, eds. (Springer-Verlag: Berlin)

\bibitem[Landolt(1992)]{landolt92}
Landolt, A. U. 1992, \aj, 104, 340 
 
\bibitem[Lupton et al.(1999)]{lupton99}
Lupton, R. H., Gunn, J. E., \& Szalay, A. S. 1999, \aj, 118, 1406

\bibitem[Lupton et al.(2001)]{lupton01}
Lupton, R., Gunn, J. E., Ivezi\'c, Z., Knapp, G. R.,
Kent, S., \& Yasuda, N. 2001, in ASP Conf. Ser. 238,
Astronomical Data Analysis Software and Systems X,
ed. F. R. Harnden, Jr., F. A. Primini, and H. E. Payne
(San Francisco: Astr. Soc. Pac.), p. 269 (astro-ph/0101420)

\bibitem[Lupton(2005)]{lupton05}
Lupton, R. 2005, \aj, submitted 

\bibitem[Matthews \& Soifer(1994)]{matthews94}
Matthews, K., \& Soifer, B.~T.\ 1994, 
ASSL Vol.~190: Astronomy with Arrays, The Next Generation, 239

\bibitem[Morgan et al.(2003)]{morgan03}
Morgan, N. D., Snyder, J. A., \& Reens, L. H. 2003, 
\aj, 126, 2145

\bibitem[Myers et al.(2003)]{myers03} 
Myers, S. T., et al. 2003, \mnras, 341, 1

\bibitem[Oguri et al.(2004)]{oguri04}
Oguri, M., et al. 2004, \pasj, 56, 399

\bibitem[Oguri et al.(2005)]{oguri05}
Oguri, M., et al. 2005, \apj, 622, 106 

\bibitem[Oscoz et al.(1997)]{oscoz97} 
Oscoz, A., Serra-Ricart, M., Mediavilla, E., Buitrago, J., 
\& Goicoechea, L. J. 1997, \aj, 491, L7

\bibitem[Peng et al.(2002)]{peng02}
Peng, C. Y., Ho, L. C., Impey, C. D., \& 
Rix, H.-W. 2002, \aj, 124, 266  

\bibitem[Pier et al.(2003)]{pier03} 
Pier, J. R., Munn, J. A., Hindsley, R. B., Hennessy, G. S., 
Kent, S. M., Lupton, R. H., \& Ivezi\'{c}, \'{Z}. 2003, 
\aj, 125, 1559

\bibitem[Pindor et al.(2004)]{pindor04}
Pindor, B., et al. 2004, \aj, 127, 1318

\bibitem[Pindor et al.(2005)]{pindor05}
Pindor, B., et al. 2005, \aj, in press (astro-ph/0509296)

\bibitem[Refsdal(1964)]{refsdal64} 
Refsdal, S. 1964, \mnras, 128, 307

\bibitem[Richards et al.(2002)]{richards02} 
Richards, G. T., et al. 2002, \aj, 123, 2945

\bibitem[Richards et al.(2004)]{richards04}
Richards, G.~T., et al.\ 2004, \apjs, 155, 257

\bibitem[Rusin et al.(2003)]{rusin03}
Rusin, D, et al.\ 2003, \apj, 587, 143

\bibitem[Schneider et al.(2005)]{schneider05}
Schneider, D. P., et al.\ 2005, \aj, 130, 367

\bibitem[Sheinis et al.(2002)]{sheinis02} 
Sheinis, A. I., Bolte, M., Epps, H. W., Kibrick, R. I., Miller, J. S., Radovan, 
M. V., Bigelow, B. C., \& Sutin, B. M. 2002, 
\pasp, 114, 851

\bibitem[Smith et al.(2002)]{smith02} 
Smith, J. A., et al. 2002, \aj, 123, 2121

\bibitem[Spergel et al.(2003)]{spergel03} 
Spergel, D. N., et al. 2003, \apjs, 148, 175

\bibitem[Stoughton et al.(2002)]{stoughton02}
Stoughton, C., et al. 2002, \aj, 123, 485

\bibitem[Strauss et al.(2002)]{strauss02} 
Strauss, M. A., et al. 2002, \aj, 124, 1810

\bibitem[Surdej et al.(1987)]{surdej87}
Surdej, J., Swings, J.-P., Magain, P., Courvoisier, T. J.-L., 
\& Borgeest, U. 1987, \nat, 329, 695

\bibitem[Sutin(1997)]{sutin97}
Sutin, B. M. 1997, Proc. SIPE, 2871, 1116

\bibitem[Tucker et al.(2005)]{tucker05} 
Tucker, D. L., et al. 2005, \aj, submitted 

\bibitem[Turner et al.(1984)]{turner84} 
Turner, E. L., Ostriker, J. P., \& Gott, J. R., III 1984, 
\apj, 284, 1

\bibitem[Turner(1990)]{turner90} 
Turner, E.~L.\ 1990, \apj, 365, L43

\bibitem[Walsh et al.(1979)]{walsh79} 
Walsh, D., Carswell, R. F., \& Weymann, R. J. 1979, \nat, 279, 381

\bibitem[Weymann et al.(1980)]{weymann80}
Weymann, R. J., et al. 1980, \nat, 285, 641

\bibitem[York et al.(2000)]{york00}
York, D. G., et al. 2000, \aj, 120, 1579

\end{thebibliography}
\end{document}